\documentclass[12pt]{article}
\usepackage{graphicx}
\usepackage{epstopdf}
\usepackage{amsmath,amssymb,amsfonts,graphicx}

\newcommand{\bpartial}{\mbox{\boldmath $\partial$}}

\newcommand{\bsigma}{\mbox{\boldmath $\sigma$}}

\begin{document}
\thispagestyle{empty}
\renewcommand{\refname}{References}

\title{\bf The Aharonov--Bohm effect in scattering theory}

\author{Yu. A. Sitenko$^1$, N. D. Vlasii$^{2}$}

\date{}

\maketitle

\begin{center}
$^{1}$Bogolyubov Institute for Theoretical Physics, National Academy
of Sciences, \\ 14-b Metrologichna
Str., Kyiv, 03680, Ukraine \\
$^{2}$Physics Department, Taras Shevchenko National University of Kyiv, \\
64 Volodymyrska str., Kyiv, 01601, Ukraine
\end{center}

\begin{abstract}
The Aharonov--Bohm effect is considered as a scattering event with
nonrelativistic charged particles of the wavelength which is less
than the transverse size of an impenetrable magnetic vortex. The
quasiclassical WKB method is shown to be efficient in solving this
scattering problem. We find that the scattering cross section
consists of two terms, one describing the classical phenomenon of
elastic reflection and another one describing the quantum phenomenon
of diffraction; the Aharonov--Bohm effect is manifested as a fringe
shift in the diffraction pattern. Both the classical and the quantum
phenomena are independent of the choice of a boundary condition at
the vortex edge, providing that probability is conserved. We show
that a propagation of charged particles can be controlled by
altering the flux of a magnetic vortex placed on their way.
\end{abstract}


\bigskip

\begin{center}
\end{center}

\bigskip
\medskip

\section{Introduction}

A comprehensive description of electromagnetism in classical
theory is given in terms of the electromagnetic field strength
acting locally and directly on charged matter. This is not the
case in quantum theory where a state of charged matter is
influenced by electromagnetic field even in the situation when the
region of the nonvanishing field strength does not overlap with
the region accessible to charged matter; the indispensable
condition is that the latter region be non-simply-connected. In
particular, a magnetic field of an infinitely long
current-carrying solenoid can be shielded from the penetration of
charged matter; such a field configuration may be denoted as an
impenetrable magnetic vortex. Perhaps, W.~Ehrenberg and
R.~E.~Siday were the first to discover theoretically the purely
quantum effect which is due to the influence of the vector
potential of the impenetrable magnetic vortex on the motion of a
charged particle, and they proposed an interference experiment to
detect this \cite{Ehre}. Ten years later the effect was
rediscovered by Y.~Aharonov and D.~Bohm who made an important
addition by predicting also an effect which is due to the
influence of the electric scalar potential \cite{Aha}.
Immediately, in a year, the first interference experiment to
verify the magnetic effect was performed \cite{Cha}, and this
caused a burst of interest opening a long way of an extensive
research both in theory and in experiment, see, e.g.,
\cite{Wu,Pes,Rui,Au,Ola,Jac,Be,Ton} and references therein; the
effect conventionally bears the name of the authors of \cite{Aha}.
Although a great progress in understanding the Aharonov--Bohm (AB)
effect has been achieved since the time of foundational works
\cite{Ehre,Aha,Cha}, one may agree with the authors of \cite{Bat}
that ``the investigation and exploitation of the AB effect remain
far from finished''.

It should be noted that the concern of paper \cite{Ehre} was in the
elaboration of the consistent theory of refraction of electron rays
in the framework of electron wave optics. Following thoroughly their
flow of reasoning, the authors arrive at the conclusion that the
medium in a spatial region outside a magnetic vortex is anisotropic
and that there exist wave-optical phenomena which are due to the
influence of the vector potential of the vortex. Finally, in the
very end of the paper, they suggest an interference experiment to
verify their findings.

An approach of paper \cite{Aha} was quite different: the authors
heuristically, with the use of few rather general and obvious formulas
(which are even left unnumbered in the text), arrive at their 
suggestion of interference experiments already in the beginning of
the paper. A powerful theoretical background behind this intuitive 
suggestion is hereafter given in Section 4. ``Exact solution for
scattering problems'' (which exceeds in its volume two preceding
sections). The background relies directly on the formalism of
quantum mechanics, namely on the Schr\"{o}dinger wave equation,
which allows the authors to consider their effect as a scattering
event. Perhaps, this was a reason, why paper \cite{Aha}, unlike
paper \cite{Ehre}, had received an immediate response from the
physics community.

In this respect it looks rather paradoxical that, whereas
interference experiments aiming at the verification of the AB effect
were numerous (see \cite{Ton}), a scattering experiment aiming at
the same was {\it never} performed. One of the motivations of the
present paper is to provide an explanation for this paradox.

Let us commence by recalling the famous formula obtained by the
authors of \cite{Aha} for the scattering differential cross section
per unit length of the magnetic vortex (see (21) and (22) in
\cite{Aha}):
\begin{equation}
\frac{\rm d\sigma^{(AB)}}{{\rm d} z\rm d\varphi}=\frac{\hbar}{2\pi p}\frac{\sin^2[e\Phi(2\hbar c)^{-1}]}
{\sin^2(\varphi/2)},\label{eq1}
\end{equation}
where $p$ is the momentum of a scattered particle, $e$ is its
charge, $\varphi$ is the scattering angle (with $\varphi=0$
corresponding to the forward direction) and $\Phi$ is the flux of
the magnetic vortex directed along the $z$-axis. A characteristic
feature of (1) is a periodic dependence in the magnetic flux with
the period equal to the London flux quantum, $2\pi\hbar c/|e|$.
Also, one could suggest that (1) describes a purely quantum effect,
since it vanishes in the formal limit of $\hbar\rightarrow 0$, or,
more physically, in the limit of large enough values of the particle
momentum, $p\rightarrow \infty$. However, result (1) corresponds to
the idealized case of the magnetic vortex of the vanishing
transverse size. Since any real magnetic vortex (either that
produced by a current-carrying solenoid, or that inside a magnet) is
of a nonzero transverse size, the quotient of squared sines in (1)
in the real case becomes a certain function of dimensionless
variable $pr_c/\hbar$, where $r_c$ is a length characterizing the
transverse size of the vortex. The transverse size effects were
taken into account in \cite{Au}, and results were found to converge
with (1) in the case of $pr_c\hbar^{-1}\rightarrow 0$. As to the
case of $pr_c\hbar^{-1}\gg 1$, it was only scarcely and
insufficiently considered, see \cite{Rui,Ola}, and, therefore, let
us dwell on this case at more length.

It should be noted in the first place that the smallest possible
values of $r_c$ can be of order 1 $\mu m$ (ferromagnetic whiskers
enclosed in screening shells) \cite{Ton}, while, in a scattering
event, the largest possible wavelengths, $\lambda=2\pi\hbar/p$, of
the lightest charged particle, electron, are of order 0.1 nm
(slowly moving electrons of energies 10-100 eV, e.g., such as
those in famous diffraction experiments by C.~J.~Davisson and
L.~H.~Germer in 1927, Nobel prize in 1937). Hence, quantity
$pr_c\hbar^{-1}$ takes values of order $10^4-10^5$ at least, and
formula (1) which might be supposed to approximate the case of
$pr_c\hbar^{-1}\ll 1$ has no relation to scattering of real
particles. Albeit this formula looks essentially quantum, the
``quantum'' factor of $\hbar/p$ appears just due to dimensional
reasons because of $r_c=0$; it is a somewhat more explicating to
write $\hbar/p$ as product $r_c\times(pr_c/\hbar)^{-1}$, where the
second (dimensionless) factor is vanishingly small for all charged
particles existing in nature. This was certainly understood by the
authors of \cite{Aha}, who never proposed cross section (1) for
the experimental verification.

It may seem quite reasonable to apply the quasiclassical (called
also semiclassical) method of G.~Wentzel, H.~A.~Kramers and
L.~Brillouin (WKB) \cite{We,Kra,Bri} to solving the problem of
scattering of real charged particles by an impenetrable magnetic
vortex: the size of the interaction region may be estimated as
being of order of $r_c$ and larger, then condition
$r_c\lambda^{-1}\gg 1$ allows one to hope that the use of the WKB
method will be adequate, see, e.g, \cite{New,Lan}. However, a
skepticism immediately arises: how can a purely and essentially
quantum effect as is the AB one be deduced with the help of a
quasiclassical (semiclassical) method which is commonly believed
to give nothing else but the yield of classical theory? In this
respect it should be noted that the findings of the authors of
\cite{Rui,Ola} indicate that scattering off an impenetrable
magnetic vortex in the case of $pr_c\hbar^{-1}\gg 1$ is reduced to
classical scattering of point particles off an impenetrable tube
with no flux.

Nevertheless, in spite of the skepticism substantiated by the
results of \cite{Rui,Ola}, we shall proceed in the present
paper by applying the WKB method to the solution of the 
scattering AB problem. Let us recall that interference 
experiments aiming at the verification of the AB effect were 
performed with rather energetic electrons. We are considering 
electrons of the same (as in the interference experiments) and 
even smaller (down to 10-20 eV) energies and study direct 
scattering of such particles on an impenetrable magnetic 
vortex, using the WKB method. Whereas the authors of 
\cite{Rui,Au,Ola} employed the Dirichlet boundary condition 
to ensure the vortex impenetrability, we go further by
employing the most general (unorthodox in the language of
\cite{Rui}) boundary condition that is required by the basic
concepts of probability conservation and self-adjointness of the
Hamilton operator. In the course of our study we shall find, in
addition to the classical effect of \cite{Rui, Ola}, a purely
quantum effect that can be denoted as the scattering AB effect
with real particles.

In the next section we consider a quantum-mechanical charged
particle in the background of an impenetrable magnetic vortex and
choose a boundary condition for the particle wave function at the
vortex edge. The WKB method is used to obtain the partial waves of
the solution to the Schr\"{o}dinger equation in Section 3. The
scattering amplitude and cross section are obtained in Section 4.
The conclusions are drawn and discussed in Section 5. We relegate
some details of calculation of the scattering amplitude to Appendix
A, while the use of the Hamilton-Jacobi equation of motion to derive
the classical effect is outlined in Appendix B.

\section{Self-adjointness of the Hamilton operator and the Robin boundary
condition}

Defining a scalar product as
$(\tilde{\chi},\chi)=\int\limits_{\Omega}{\rm
d}^3x\tilde{\chi}^*\chi$, we get, using integration by parts,

\begin{eqnarray}
(\tilde{\chi},H\chi)\equiv \int\limits_{\Omega}{\rm d}^3x\tilde{\chi}^*\left\{\left[
-\frac{1}{2m}\left(\hbar{\bpartial}-{\rm i}\frac{e}{c}{\bf A}\right)^2+V\right]\chi\right\}=
\nonumber \\
=\!\int\limits_{\Omega}\!{\rm d}^3x\left\{\left[
-\frac{1}{2m}\left(\hbar{\bpartial}-{\rm i}\frac{e}{c}{\bf A}\right)^2\!+\!V\right]\tilde{\chi}\right\}^*\chi-
\frac{\hbar}{2m}\int\limits_{\partial\Omega}\!{\rm d}{\bsigma}\cdot
\left\{\tilde{\chi}^*\left[\left(\hbar{\bpartial}-{\rm i}\frac{e}{c}{\bf{A}}\right)\chi\right]\right.\!-\nonumber \\
\left.-\left[\left(\hbar{\bpartial}+{\rm i}\frac{e}{c}{\bf{A}}\right)\tilde{\chi}^*\right]\chi\right\}\equiv
\left(H^\dag\tilde{\chi},\chi\right)-{\rm i}\hbar\int\limits_{\partial\Omega}{\rm d}{\bsigma}\cdot{\bf u},
\label{eq2}
\end{eqnarray}
where $\partial\Omega$ is a two-dimensional surface bounding the
three-dimensional spatial region $\Omega$,
\begin{equation}
H=H^\dag=-\frac{1}{2m}\left(\hbar\bpartial-{\rm i}\frac{e}{c}{\bf A}\right)^2+V\label{eq3}
\end{equation}
is the formal expression for a general quantum-mechanical Hamilton
operator in an external electromagnetic field (electric scalar
potential is included in the general potential energy part $V$), and
\begin{equation}
{\bf u}=-\frac{\rm i}{2m}\left\{\tilde{\chi}^*\left[\left(\hbar\bpartial-{\rm i}\frac{e}{c}{\bf A}\right)\chi\right]
-\left[\left(\hbar\bpartial+{\rm i}\frac{e}{c}{\bf A}\right)\tilde{\chi}^*\right]\chi\right\}.\label{eq4}
\end{equation}
Operator $H$ is Hermitian (symmetric),
\begin{equation}
(\tilde{\chi},H\chi)=(H^\dag\tilde{\chi},\chi),\label{eq5}
\end{equation}
if
\begin{equation}
\int\limits_{\partial\Omega}{\rm d}\bsigma\cdot{\bf u}=0.\label{eq6}
\end{equation}
The latter condition can be satisfied in various ways by imposing
different boundary conditions for $\chi$ and $\tilde{\chi}$.
However, among the whole variety, there may exist a possibility that
a boundary condition for $\tilde{\chi}$ is the same as that for
$\chi$; then the domain of definition of $H^\dag$ (set of functions
$\tilde{\chi}$) coincides with that of $H$ (set of functions
$\chi$), and operator $H$ is called self-adjoint. Whether such a
possibility exists is in general determined by the Weyl -- von
Neumann theory of self-adjoint operators, see, e.g., \cite{Ree}. The
action of a self-adjoint operator results in functions belonging to
its domain of definition only, and, therefore, a multiple action and
functions of such an operator (for instance, the evolution operator)
can be consistently defined.

In the present case, the problem of self-adjointness of operator $H$
(3) is resolved by imposing a boundary condition in the form
\begin{eqnarray}
\biggl.\left[\sin(\rho\pi)\,{\bf n}\cdot\left(\bpartial-\frac{{\rm i}e}{\hbar c}{\bf A}\right)\chi+
\cos(\rho\pi)\frac{1}{\bf n\cdot{\bf x}}\chi\right]\biggr|_{{\bf x}\in\partial\Omega}=
\nonumber \\
\biggl.=\left[\sin(\rho\pi)\,{\bf n}\cdot\left(\bpartial-\frac{{\rm i}e}{\hbar c}{\bf A}\right)\tilde{\chi}+
\cos(\rho\pi)\frac{1}{\bf n\cdot{\bf x}}\tilde{\chi}\right]\biggr|_{{\bf x}\in\partial\Omega}=0,
\label{eq7}
\end{eqnarray}
where ${\bf n}$ is the internal unit normal to boundary
$\partial\Omega$ and $\rho$ is the real parameter (the self-adjoint
extension parameter). Defining current
\begin{equation}
{\bf j}=-\frac{\rm i}{2m}\left\{\chi^*\left[\left(\hbar\bpartial-{\rm i}\frac{e}{c}{\bf A}\right)\chi\right]-
\left[\left(\hbar\bpartial+{\rm i}\frac{e}{c}{\bf A}\right)\chi^*\right]\chi\right\},\label{eq8}
\end{equation}
we note that self-adjointness condition (7) results in the vanishing
of the normal component of the current at the boundary:
\begin{equation}
{\bf n}\cdot{\bf j}|_{{\bf x}\in \partial\Omega}=0.\label{eq9}
\end{equation}

If $\chi$ is a solution to the Schr\"{o}dinger equation,
\begin{equation}
{\rm i}\hbar\frac{\partial}{\partial t}\chi(t,{\bf x})=H\chi(t,{\bf x}),\label{eq10}
\end{equation}
then current (8) is the probability current density satisfying the
continuity equation
\begin{equation}
\frac{\partial}{\partial t}|\chi^2|+\bpartial\cdot{\bf j}=0.\label{eq11}
\end{equation}
Thus, self-adjointness of the Hamilton operator ensures probability
conservation, $\frac{\partial}{\partial
t}\int\limits_{\Omega}d^3x|\chi|^2=0$.

In the following we consider a quantum-mechanical charged particle
in the background of a static magnetic field, hence $V=0$. The
magnetic field is confined to an infinite tube which is classically
impenetrable to the charged particle; thus, spatial region $\Omega$
is an infinite three-dimensional space with the exclusion of the tube
containing the magnetic flux lines, and boundary surface
$\partial\Omega$ is a surface of the tube. Assuming the cylindrical
symmetry of the flux tube, we use cylindrical coordinates ${\bf
x}=(r,\varphi,z)$ with the $z$-axis coinciding with the axis of the
tube and choose the components of the vector potential in $\Omega$
in the form
\begin{equation}
A_r=A_z=0,\qquad A_\varphi=\frac{\Phi}{2\pi},\label{eq12}
\end{equation}
where $\Phi$ is the total magnetic flux confined in the tube. Then
the Hamilton operator takes form
\begin{equation}
H=-\frac{\hbar^2}{2m}\left[\frac{1}{r}\partial_rr\partial_r+
\frac{1}{r^2}\left(\partial_\varphi-\frac{{\rm i}e\Phi}{2\pi\hbar c}\right)^2+\partial_z^2\right],\label{eq13}
\end{equation}
and the solution to Schr\"{o}dinger equation (10) with $H$ (13) is
\begin{equation}
\chi(t,{\bf x})=\exp(-{\rm i}Et\hbar^{-1}+{\rm i}p_zz\hbar^{-1})\psi(r,\varphi),\label{eq14}
\end{equation}
where $\psi(r,\varphi)$ is the solution to equation
\begin{equation}
\left[\frac{1}{r}\partial_rr\partial_r+\frac{1}{r^2}\left(\partial_\varphi-\frac{{\rm i}e\Phi}{2\pi\hbar c}\right)^2+
\frac{p^2}{\hbar^2}\right]\psi(r,\varphi)=0,\label{eq15}
\end{equation}
and obeys the boundary condition, see (7),
\begin{equation}
[\sin(\rho\pi)r\partial_r\psi(r,\varphi)+\cos(\rho\pi)\psi(r,\varphi)]|_{r=r_c}=0;\label{eq16}
\end{equation}
$r_c$ is the tube radius, $E=(p_z^2+p^2)(2m)^{-1}$ is the energy of
the stationary scattering state, $p_z$ and $p$ are the momenta in
the longitudinal and radial transverse directions, respectively.
One could recognize that (16) is known as the Robin boundary
condition; the case of $\rho=0$ corresponds to the Dirichlet
condition (perfect reflectivity of the boundary) and the case of
$\rho=1/2$ corresponds to the Neumann condition (absolute rigidity
of the boundary). It should be emphasized that parameter $\rho$ is
in general dependent on $z$ and $\varphi$. Thus, the ``number'' of
self-adjoint extension parameters is infinite, moreover, it is not
countable but is of power of a continuum. This distinguishes the
case of an extended boundary from the case of an excluded point
(contact interaction) when the number of self-adjoint extension
parameters is finite, being equal to $n^2$ for the deficiency index
equal to \{$n,n$\} (see, e.g., \cite{Alb}).

The motion of the particle in the longitudinal direction is free,
and our aim is to determine the particle motion in the orthogonal
plane, i.e. to find a solution to (15) and (16).

\section{WKB method}

Let us start with the decomposition of wave function
$\psi(r,\varphi)$ into partial waves
\begin{equation}
\psi(r,\varphi)=\sum\limits_{n\in \mathbb{Z}}e^{{\rm i}n\varphi}a_nR_n(r),\label{eq17}
\end{equation}
where $\mathbb{Z}$ is the set of integer numbers. The radial
component of each partial wave satisfies equation
\begin{equation}
[\Delta_r+P_n^2(r)\hbar^{-2}]R_n(r)=0,\label{eq18}
\end{equation}
where $\Delta_r=r^{-1}\partial_rr\partial_r$ and
\begin{equation}
P_n(r)=\sqrt{p^2-\hbar^2(n-\mu)^2r^{-2}},\label{eq19}
\end{equation}
$\mu=e\Phi(2\pi\hbar c)^{-1}$. Note that the amplitude of
$\psi(r,\varphi)$ (17) is periodic in the value of flux $\Phi$
with the period equal to the London flux quantum.

We present the radial component of the partial wave as
$R_n(r)=\exp\left[\frac{\rm i}{\hbar}\Sigma_n(r)\right]$, where
$\Sigma_n(r)$ satisfies the Ricatti-type equation
\begin{equation}
\frac{\rm i}{\hbar}\Delta_r\Sigma_n-\frac{1}{\hbar^2}(\partial_r\Sigma_n)^2+\frac{1}{\hbar^2}P_n^2=0.\label{eq20}
\end{equation}
Expanding $\Sigma_n(r)$ into series
\begin{equation}
\Sigma_n(r)=\sum\limits_{l=0}^{\infty}\left(\frac{\hbar}{{\rm i}pr_c}\right)^l\Sigma_n^{(l)}(r)\label{eq21}
\end{equation}
and collecting terms of the same order in $\hbar$, we obtain the
system of equations:
\begin{equation}
\begin{array}{l}
  \left[\partial_r\Sigma_n^{(0)}\right]^2=P_n^2, \\
  2\left[\partial_r\Sigma_n^{(0)}\right]\left[\partial_r\Sigma_n^{(1)}\right]=-pr_c\Delta_r\Sigma_n^{(0)}, \\
  2\left[\partial_r\Sigma_n^{(0)}\right]\left[\partial_r\Sigma_n^{(2)}\right]=-pr_c\Delta_r\Sigma_n^{(1)}-
\left[\partial_r\Sigma_n^{(1)}\right]^2, \\
  \qquad\qquad \vdots \\
   2\left[\partial_r\Sigma_n^{(0)}\right]\left[\partial_r\Sigma_n^{(2l)}\right]=-pr_c\Delta_r\Sigma_n^{(2l-1)}-
2\sum\limits_{l'=1}^{l-1}\left[\partial_r\Sigma_n^{(l')}\right]\left[\partial_r\Sigma_n^{(2l-l')}\right]-
\left[\partial_r\Sigma_n^{(l)}\right]^2,\\
  2\left[\partial_r\Sigma_n^{(0)}\right]\left[\partial_r\Sigma_n^{(2l+1)}\right]=-pr_c\Delta_r\Sigma_n^{(2l)}-
2\sum\limits_{l'=1}^{l}\left[\partial_r\Sigma_n^{(l')}\right]\left[\partial_r\Sigma_n^{(2l+1-l')}\right], \\
   \qquad\qquad \vdots
  \end{array} \label{eq22}
\end{equation}
Consecutively solving equations beginning from the top, one can
obtain $\Sigma_n(r)$ (21) up to an arbitrary order in expansion
parameter $\hbar/(pr_c)$. The terms corresponding to odd $l$
contribute to the amplitude of $R_n(r)$, while the terms
corresponding to even $l$ contribute to the phase of $R_n(r)$.

However, the power of the WKB method reveals fully itself in those
cases when it suffices to take account for the first two terms,
$\Sigma_n^{(0)}(r)$ and $\Sigma_n^{(1)}(r)$, only. It should be
noted that equation (18), or (20), has its analogue in optics,
describing the propagation of waves in media \cite{Wolf}: quotient
$P_n(r)/p$ corresponds to the index of refraction. Terms
$\Sigma_n^{(l)}(r)$ $(l\geq 2)$ are negligible, if the variation of
the refraction index along the ray of propagation becomes
appreciable on the distances which are much larger than the
effective wavelength,
\begin{equation}
|P_n^{-1}\,\bpartial\cdot({\bf n}P_n)|^{-1}\gg \hbar|P_n|^{-1},\label{eq23}
\end{equation}
where ${\bf n}$ points in the direction of the ray which in our
case is orthogonal to the boundary. In view of (19) this yields
condition
\begin{equation}
\frac{\hbar|p^2r^2|}{|p^2r^2-\hbar^2(n-\mu)^2|^{3/2}}\ll 1,\label{eq24}
\end{equation}
which is satisfied in the quasiclassical region,
\begin{equation}
pr\gg\hbar|n-\mu|,\qquad pr\gg\hbar,\label{eq25}
\end{equation}
as well as in the deeply nonclassical region
\begin{equation}
pr\ll\hbar|n-\mu|,\qquad |n-\mu|\gg 1.\label{eq26}
\end{equation}
Solving the first two equations in (22), we get
\begin{equation}
\Sigma_n^{(0)}(r)=\pm\int\limits_{}^{r}{\rm d}rP_n(r),\qquad
\Sigma_n^{(1)}(r)=-\frac{1}{2}pr_c\ln[rP_n(r)].\label{eq27}
\end{equation}
It should be noted that the second equation in (22) is nothing more
but the continuity equation, $\bpartial\cdot{\bf j}_n$, for the
partial wave current,
$$
{\bf j}_n=-\frac{{\rm i}\hbar}{2m}\left[e^{-{\rm i}n\varphi}R_n^*(\bpartial e^{{\rm i}n\varphi}R_n)-
(\bpartial e^{-{\rm i}n\varphi}R_n^*)e^{{\rm i}n\varphi}R_n\right],
$$
in the quasiclassical region.

Actually, the WKB approximation (i.e. the neglect of
$\Sigma_n^{(l)}(r)$ with $l\geq 2$) is efficient in a much more
extended regions than those given by (25) and (26). The WKB
approximation is not valid in the vicinity of $r=r_t$ which is
defined by
\begin{equation}
P_n(r_t)=0;\label{eq28}
\end{equation}
this point is the turning point of a classical trajectory of the
particle. In view of (27), the two linearly independent solutions to
(18) in the quasiclassical region are
\begin{equation}
R_{n,{\rm out}}^{(\pm)}(r)=(rP_n)^{-1/2}\exp\left(\pm\frac{\rm i}{\hbar}\int\limits_{r_t}^{r}{\rm d}r\,P_n\right),
\qquad r>r_t,\label{eq29}
\end{equation}
while the two linearly independent solutions to (18) in the deeply
nonclassical region are
\begin{equation}
R_{n,{\rm in}}^{(\pm)}(r)=(r\Pi_n)^{-1/2}\exp\left(\pm\frac{1}{\hbar}\int\limits_{r_t}^{r}{\rm d}r\,\Pi_n\right),
\qquad r<r_t,\label{eq30}
\end{equation}
where
\begin{equation}
\Pi_n(r)=\sqrt{\hbar^2(n-\mu)^2r^{-2}-p^2}.\label{eq31}
\end{equation}
The crucial question is how oscillating solutions (29) in the outer
region go over to solutions (30) in the inner region and vice verse.
To answer this question, one has to find a solution in the vicinity
of the turning point, where $P_n(r)=p\sqrt{\frac{2}{r_t}(r-r_t)}$
and $\Pi_n(r)=p\sqrt{\frac{2}{r_t}(r_t-r)}$, and to match it with
the solutions in the inner and the outer regions. But a more
instructive way, as stated in \cite{Lan}, is to consider formally
$R_n(r)$ as a function of complex variable $r$ and to perform
transitions along paths in the complex plane, where condition (24)
is satisfied. Namely, a path goes along the real positive semiaxis
up to the left vicinity of the turning point, circumvents it along a
semicircle in the upper or lower half-plane, and then goes again
along the real positive semiaxis. For the transition from the inner
to the outer region we have
\begin{equation}
\Pi_n(r)\rightarrow P_n(r)e^{\pm {\rm i}\pi/2},\label{eq32}
\end{equation}
where the ($\pm$) sign corresponds to the path in the upper or lower
half-plane, and
\begin{equation}
R_{n,{\rm in}}^{(+)}(r)\rightarrow R_{n,{\rm out}}^{(+)}(r)e^{-{\rm i}\pi/4}+R_{n,{\rm out}}^{(-)}
(r)e^{{\rm i}\pi/4}.\label{eq33}
\end{equation}
$R_{n,{\rm in}}^{(-)}(r)$, in contrast to $R_{n,{\rm in}}^{(+)}(r)$,
is divergent in the inner region. For the transition from the outer
to the inner region we have
\begin{equation}
P_n(r)\rightarrow \Pi_n(r)e^{\pm {\rm i}\pi/2},\label{eq34}
\end{equation}
and
\begin{equation}
\frac{\rm i}{2}\left[R_{n,{\rm out}}^{(+)}(r)e^{-{\rm i}\pi/4}-
R_{n,{\rm out}}^{(-)}(r)e^{{\rm i}\pi/4}\right]
\rightarrow R_{n,{\rm in}}^{(-)}(r),\label{eq35}
\end{equation}
where the account is taken for the fact that $R_{n, {\rm
in}}^{(-)}(r)$ is real.

Boundary condition (16) in terms of partial waves takes form
\begin{equation}
[\sin(\rho\pi)r\partial_rR_n(r)+\cos(\rho\pi)R_n(r)]|_{r=r_c}=0.\label{eq36}
\end{equation}
In the case $r_t<r_c\leq r$ we choose
\begin{equation}
R_n(r)=\left[R_{n,{\rm out}}^{(-)}(r)-C_n(r_c,\rho)R_{n,{\rm out}}^{(+)}
(r)\right]e^{{\rm i}\pi/4},\label{eq37}
\end{equation}
where the factor of $\exp({\rm i}\pi/4)$ is inserted for future
convenience. Coefficient $C_n(r_c,\rho)$ is determined from
condition (36):
\begin{eqnarray}
C_n(r_c,\rho)=\frac{R_{n,{\rm out}}^{(-)}(r_c)}{R_{n,{\rm out}}^{(+)}(r_c)}
\,\frac{\cot(\rho\pi)+[r\partial_r\ln R_{n,{\rm out}}^{(-)}(r)]|_{r=r_c}}
{\cot(\rho\pi)+[r\partial_r\ln R_{n,{\rm out}}^{(+)}(r)]|_{r=r_c}}=
\nonumber \\
=\exp\left\{-\frac{2{\rm i}}{\hbar}\int\limits_{r_t}^{r_c}{\rm d}r\,P_n-2{\rm i}\arctan\left[
\frac{r_cP_n(r_c)\hbar^{-1}}{\cot(\rho\pi)-\frac{1}{2}\,\frac{p^2}{P_n^2(r_c)}}\right]\right\},
\qquad r_c>r_t;
\label{eq38}
\end{eqnarray}
note that the explicit form of the turning point (i.e. the solution
to (28)) is evidently $r_t=\hbar|n-\mu|/p$. In the case $r\leq r_c
<r_t$ we choose
\begin{equation}
R_n(r)=R_{n,{\rm in}}^{(+)}(r)-C_n(r_c,\rho)\left[R_{n,{\rm in}}^{(-)}(r)+cR^{(+)}_{n,{\rm in}}(r)\right],\label{eq39}
\end{equation}
where coefficient $C_n(r_c,\rho)$ is determined from condition (36):
$$
C_n(r_c,\rho)=\frac{R_{n,{\rm in}}^{(+)}(r_c)}{R_{n,{\rm in}}^{(-)}(r_c)+cR_{n,{\rm in}}^{(+)}(r_c)}
\times
$$
$$
\times\frac{\cot(\rho\pi)+\left.\left[r\partial_r\ln R_{n,{\rm
in}}^{(+)}(r)\right]\right|_{r=r_c}}
{\cot(\rho\pi)+\left.\left\{r\partial_r\ln \left[R_{n,{\rm
in}}^{(-)}(r)+cR_{n,{\rm in}}^{(+)}(r)\right]\right\}
\right|_{r=r_c}}, \,\,r_c<r_t,
$$
and constant $c$ is to be fixed before long. Using correspondence
rules (33) and (35), we continue $R_n(r)$ (39) to the case
$r_c<r_t<r$ as
$$
R_n(r)=R_{n,{\rm out}}^{(+)}(r)e^{-{\rm i}\pi/4}+R_{n,{\rm out}}^{(-)}(r)e^{{\rm i}\pi/4}-
$$
$$
-C_n(r_c,\rho)\left\{\frac{\rm i}{2}\left[R_{n,{\rm out}}^{(+)}(r)e^{-{\rm i}\pi/4}-
R_{n,{\rm out}}^{(-)}(r)e^{{\rm i}\pi/4}\right]+
c\left[R_{n,{\rm out}}^{(+)}(r)e^{-{\rm i}\pi/4}+
R_{n,{\rm out}}^{(-)}(r)e^{{\rm i}\pi/4}\right]\right\}.
$$
The converging wave, $R_{n,{\rm out}}^{(-)}(r)$, should disappear
from the part with coefficient $C_n(r_c,\rho)$, cf. (37). This fixes
the constant, $c={\rm i}/2$, yielding in the case $r_c<r_t<r$
\begin{equation}
R_n(r)=R_{n,{\rm out}}^{(+)}(r)e^{-{\rm i}\pi/4}+R_{n,{\rm out}}^{(-)}(r)e^{{\rm i}\pi/4}-
C_n(r_c,\rho)R_{n,{\rm out}}^{(+)}(r)e^{{\rm i}\pi/4},\label{eq40}
\end{equation}
with
\begin{eqnarray}
C_n(r_c,\rho)=\exp\left(\frac{2}{\hbar}\int\limits_{r_t}^{r_c}{\rm d}r\Pi_n\right)\times\nonumber \\
\times\left[\frac{{\rm cot}(\rho\pi)+\frac 12\,\frac{p^2}{\Pi_n^2(r_c)}-r_c\Pi_n(r_c)\hbar^{-1}}
{{\rm cot}(\rho\pi)+\frac 12\,\frac{p^2}{\Pi_n^2(r_c)}+r_c\Pi_n(r_c)\hbar^{-1}}+
\frac{\rm i}{2}\exp\left(\frac{2}{\hbar}\int\limits_{r_t}^{r_c}{\rm d}r\Pi_n\right)\right]^{-1},\,\,r_c<r_t.\label{eq41}
\end{eqnarray}
We can rewrite (37) in the case $r_t<r_c\leq r$ as
\begin{equation}
R_n(r)=R_{n,{\rm out}}^{(+)}(r)e^{-{\rm i}\pi/4}+R_{n,{\rm out}}^{(-)}(r)e^{{\rm i}\pi/4}-
\left[e^{-{\rm i}\pi/2}+C_n(r_c,\rho)\right]R_{n,{\rm out}}^{(+)}(r)e^{{\rm i}\pi/4}.\label{eq42}
\end{equation}
We conclude this section by stating that (40) and (42) give the
radial components of partial waves of $\psi(r,\varphi)$ (17) in the
quasiclassical region in the WKB approximation.

\section{Scattering amplitude and cross section}

Going over to asymptotics $pr\hbar^{-1}\rightarrow\infty$, where
\begin{equation}
rP_n(r)=pr+O[\hbar^2(pr)^{-1}],\qquad \int\limits_{r_t}^{r}drP_n=pr-\frac{\hbar}{2}
|n-\mu|\pi+O[\hbar^2(pr)^{-1}],\label{eq43}
\end{equation}
we obtain
\begin{equation}
\psi(r,\varphi)=\psi^{(0)}(r,\varphi)+\psi^{(c)}(r,\varphi),\label{eq44}
\end{equation}
where
\begin{equation}
\psi^{(0)}(r,\varphi)=\frac{2}{\sqrt{pr}}\sum\limits_{n\in \mathbb{Z}}e^{{\rm i}n\varphi}
a_n\cos\left(pr\hbar^{-1}-\frac 12|n-\mu|\pi-\frac 14\pi\right),\label{eq45}
\end{equation}
and
\begin{eqnarray}
\psi^{(c)}(r,\varphi)=-\frac{e^{{\rm i}(pr\hbar^{-1}+\pi/4)}}{\sqrt{pr}}\left\{\sum\limits_{|n-\mu|\leq pr_c/\hbar}
e^{{\rm i}n\varphi}a_ne^{-\frac{\rm i}{2}|n-\mu|\pi}\left[e^{-{\rm i}\pi/2}+C_n(r_c,\rho)\right]+\right.
\nonumber \\
\left.+\sum\limits_{|n-\mu|>pr_c/\hbar}e^{{\rm i}n\varphi}a_ne^{-\frac{\rm i}{2}|n-\mu|\pi}
C_n(r_c,\rho)\right\}.\label{eq46}
\end{eqnarray}
To determine coefficient $a_n$, let us recall the asymptotics of the
partial wave decomposition of a plane wave,
\begin{eqnarray}
e^{{\rm i}pr\hbar^{-1}\cos \varphi}=\frac{1}{\sqrt{2\pi pr\hbar^{-1}}}\sum\limits_{n\in \mathbb{Z}}
e^{{\rm i}n\varphi}\left[e^{{\rm i}(pr\hbar^{-1}-\pi/4)}+e^{{\rm i}|n|\pi}e^{-{\rm i}(pr\hbar^{-1}-\pi/4)}\right]=
\nonumber \\
=\sqrt{\frac{2\pi\hbar}{pr}}\left[\Delta(\varphi)e^{{\rm i}(pr\hbar^{-1}-\pi/4)}
+\Delta(\varphi-\pi)e^{-({\rm i}pr\hbar^{-1}-\pi/4)}\right],\label{eq47}
\end{eqnarray}
where
$\Delta(\varphi)=(2\pi)^{-1}\sum\limits_{n\in\mathbb{Z}}e^{{\rm
i}n\varphi}$ is the angular delta-function,
$\Delta(\varphi+2\pi)=\Delta(\varphi)$. The asymptotics of the plane
wave is naturally interpreted as a superposition of two cylindrical
waves: the diverging one, $e^{{\rm i}pr\hbar^{-1}}/\sqrt{r}$, going
in the forward, $\varphi=0$, direction and the converging one,
$e^{-{\rm i}pr\hbar^{-1}}/\sqrt{r}$, coming from the backward,
$\varphi=\pi$, direction. Namely the converging cylindrical wave
should be present without distortions in the asymptotics of wave
function $\psi(r,\varphi)$ (44), whereas the diverging cylindrical
wave is distorted and differs from that in (47); actually, this is
the condition that $\psi(r,\varphi)$ be the scattering state
solution. Equating terms before $e^{-{\rm i}pr\hbar^{-1}}/\sqrt{r}$
in $\psi(r,\varphi)$ to the terms before $e^{-{\rm
i}pr\hbar^{-1}}/\sqrt{r}$ in the first line of (47), we get
\begin{equation}
a_n=\frac{1}{\sqrt{2\pi\hbar^{-1}}}\exp\left[{\rm i}\left(|n|-\frac 12|n-\mu|\right)\pi\right].\label{eq48}
\end{equation}
As a result, we obtain the following expression for the
$r_c$-independent part of the wave function
\begin{equation}
\psi^{(0)}(r,\varphi)=\psi_0^{(0)}(r,\varphi)+f_0(p,\varphi)\frac{e^{{\rm i}(pr\hbar^{-1}+\pi/4)}}{\sqrt{r}},\label{eq49}
\end{equation}
where
\begin{eqnarray}
\psi_0^{(0)}(r,\varphi)=e^{{\rm i}pr\hbar^{-1}\cos \varphi}e^{{\rm i}\mu[\varphi-{\rm sgn}(\varphi)\pi]}=\nonumber \\
=\sqrt{\frac{2\pi\hbar}{pr}}\left[e^{-{\rm i}\mu{\rm sgn}(\varphi)\pi}
\Delta(\varphi)e^{{\rm i}(pr\hbar^{-1}-\pi/4)}+\Delta(\varphi-\pi)
e^{-{\rm i}(pr\hbar^{-1}-\pi/4)}\right],\label{eq50}
\end{eqnarray}
is the incident wave and
\begin{equation}
f_0(p,\varphi)=\frac{{\rm sin}(\mu\pi)}{\sqrt{2\pi p\hbar^{-1}}}
\sum\limits_{n\in\mathbb{Z}}{\rm sgn}(n-\mu)e^{{\rm i}n\varphi}=
{\rm i}\frac{\sin(\mu\pi)}{\sqrt{2\pi p\hbar^{-1}}}\frac{e^{{\rm i}\left([\![\mu]\!]+\frac 12\right)\varphi}}
{\sin(\varphi/2)}\label{eq51}
\end{equation}
is the scattering amplitude \cite{Aha}; the sign function is ${\rm
sgn}(u)=\pm1$ at $u\gtrless 0$, $[\![u]\!]$ denotes the integer part
of quantity $u$ (i.e. the integer which is less than or equal to
$u$) and it is implied that $-\pi<\varphi<\pi$. The squared absolute
value of (51) yields differential cross section (1). Incident wave
(50) differs from the plane wave: the distortions are due to the
long-range nature of the vector potential outside the flux tube. The
apparent divergence of amplitude (51) in the forward direction is a
phantom, because (51) is valid under condition
$\sqrt{pr/\hbar}|\sin(\varphi/2)|\gg1$, whereas, otherwise, the
divergence is absent and the discontinuity in incident wave (50) at
$\varphi\rightarrow \pm0$ is cancelled, yielding
$\psi^{(0)}(r,\varphi)$ which is continuous and differentiable at
$\varphi=0$, see, e.g., \cite{SiV,Vl} and references therein.
Amplitude $f_0$ (51) is of order
$\sqrt{r_c}O\left[\sqrt{\hbar/(pr_c)}\right]$ and is negligible as
compared to the contribution from the $r_c$-dependent part of the
wave function, which can be presented in the following form:
\begin{equation}
\psi^{(c)}(r,\varphi)=[f_1(p,\varphi)+f_2(p,\varphi)+f_3(p,\varphi)]
\frac{e^{{\rm i}(pr\hbar^{-1}+\pi/4)}}{\sqrt{r}},\label{eq52}
\end{equation}
where
\begin{equation}
f_1(p,\varphi)=\frac{\rm i}{\sqrt{2\pi p\hbar^{-1}}}\sum\limits_{|n-\mu|\leq pr_c/\hbar}
e^{{\rm i}n\varphi}e^{{\rm i}(|n|-|n-\mu|)\pi},\label{eq53}
\end{equation}
\begin{equation}
f_2(p,\varphi)=-\frac{1}{\sqrt{2\pi p\hbar^{-1}}}\sum\limits_{|n-\mu|\leq pr_c/\hbar}
e^{{\rm i}n\varphi}e^{{\rm i}(|n|-|n-\mu|)\pi}\,C_n(r_c,\rho),\label{eq54}
\end{equation}
\begin{equation}
f_3(p,\varphi)=-\frac{1}{\sqrt{2\pi p\hbar^{-1}}}\sum\limits_{|n-\mu|> pr_c/\hbar}
e^{{\rm i}n\varphi}e^{{\rm i}(|n|-|n-\mu|)\pi}\,C_n(r_c,\rho).\label{eq55}
\end{equation}
We show in Appendix A (see (A.22)) that amplitude $f_3$ (55),
although exceeding amplitude $f_0$ (51), is still negligible,
being of order $\sqrt{r_c}\,O[\hbar^{1/6}(pr_c)^{-1/6}]$.
Amplitude $f_2$ (54) is of nonnegligible value which is calculated
in Appendix A:
\begin{eqnarray}
f_2(p,\varphi)=-\sqrt{\frac{r_c}{2}|\sin(\varphi/2)|}\exp\left\{-2{\rm i}pr_c\hbar^{-1}|\sin(\varphi/2)|+
{\rm i}\mu[\varphi-{\rm sgn}(\varphi)\pi]-{\rm i}\pi/4\right\}\times\nonumber \\
\times\exp\left\{-2{\rm i}\arctan\left[\frac{2pr_c\hbar^{-1}|\sin^3(\varphi/2)|}
{2\cot(\rho\pi)\sin^2(\varphi/2)-1}\right]\right\}.\label{eq56}
\end{eqnarray}
Amplitude $f_1$ (53) is a finite sum of geometric progression, which
is straightforwardly calculated:
\begin{equation}
f_1(p,\varphi)={\rm i}\sqrt{\frac{2\hbar}{\pi p}}e^{{\rm i}([\![\mu]\!]+\frac 12)\varphi}
\frac{\sin(s_c\varphi/2)}{\sin(\varphi/2)}\cos(\mu\pi+s_c\varphi/2)\label{eq57}
\end{equation}
in the case
\begin{equation}
[\![pr_c\hbar^{-1}+\mu]\!]-[\![\mu]\!]=[\![pr_c\hbar^{-1}-\mu]\!]+[\![\mu]\!]+1=s_c,\label{eq58}
\end{equation}
or
\begin{eqnarray}
f_1(p,\varphi)={\rm i}\sqrt{\frac{2\hbar}{\pi p}}e^{{\rm i}([\![\mu]\!]+\frac 12\mp\frac 12)\varphi}
\left\{\frac{\sin[(s_c+1/2)\varphi/2]}{\sin(\varphi/2)}\cos\left[\mu\pi+(s_c+1/2)\varphi/2\right]-\right.
\nonumber \\\left.-\tan(\varphi/4)\pm{\rm i}\right\}\label{eq59}
\end{eqnarray}
in the case
\begin{equation}
[\![pr_c\hbar^{-1}+\mu]\!]-[\![\mu]\!]-\frac 12\pm\frac 12=[\![pr_c\hbar^{-1}-\mu]\!]+[\![\mu]\!]+
\frac 12\mp\frac 12=s_c.\label{eq60}
\end{equation}
Since $s_c\approx pr_c\hbar^{-1}\gg 1$, amplitude $f_1$ is strongly
peaked in the forward direction (actually as a smoothed
delta-function), whereas amplitude $f_2$ is vanishing in the forward
direction. Therefore, the interference between the amplitudes,
$f_1^*f_2+f_2^*f_1$, is vanishing in the same manner as
$\sqrt{|\varphi|}\Delta(\varphi)=0$. The differential cross section
of the scattering process, ${\rm d}\sigma/({\rm
d}z{\rm d}\varphi)$, is hence a sum of two terms:
\begin{equation}
\frac{\rm d\sigma_2}{{\rm d} z \rm d\varphi}\equiv|f_2(p,\varphi)|^2=\frac{r_c}{2}\left|\sin\frac{\varphi}{2}\right|\label{eq60}
\end{equation}
and
\begin{equation}
\frac{\rm d\sigma_1}{{\rm d} z \rm d\varphi}\equiv|f_1(p,\varphi)|^2=4r_c\Delta_{\frac{pr_c}{2\hbar}}(\varphi)
\cos^2\left[\left(pr_c\varphi+\frac{e}{c}\Phi\right)(2\hbar)^{-1}\right],\label{eq62}
\end{equation}
where we have recalled that $\mu=e\Phi(2\pi\hbar c)^{-1}$ and
introduced function
\begin{equation}
\Delta_y(\varphi)=\frac{1}{4\pi
y}\frac{\sin^2(y\varphi)}{\sin^2(\varphi/2)}\qquad
(-\pi<\varphi<\pi),\label{eq63}
\end{equation}
which at $y\gg1$ can be regarded as a regularized (smoothed)
delta-function,
$$
\lim\limits_{y\rightarrow\infty}\Delta_y(\varphi)=\Delta(\varphi),\qquad
\Delta_y(0)=\frac{y}{\pi},\quad\int\limits_{-\pi}^{\pi}{\rm
d}\varphi\Delta_y(\varphi)=1+O(y^{-2}).
$$

Term (61) which is independent of the self-adjoint extension
parameter, the magnetic flux and even the particle momentum is
obtainable in the framework of classical theory, for instance, with
the use of the Hamilton-Jacobi equation of motion, see Appendix B.
Amplitude $f_2$ (54) can be rewritten as
\begin{equation}
f_2(p,\varphi)=-\frac{\rm i}{\sqrt{2\pi p\hbar^{-1}}}
\sum\limits_{|n-\mu|\leq pr_c/\hbar}\exp\left\{{\rm i}\left[n\varphi+2\delta_n^{\rm WKB}(pr_c\hbar^{-1})-
2\omega_n(pr_c\hbar^{-1},\rho)\right]\right\},\label{eq64}
\end{equation}
where
\begin{eqnarray}
\delta_n^{\rm WKB}(pr_c\hbar^{-1})=\frac{1}{2}n{\rm sgn}(n-\mu)\pi-\frac{1}{\hbar}
\int\limits_{\infty}^{r_c}{\rm d}r(P_n-p)-\frac{1}{\hbar}pr_c-\frac 14\pi=\nonumber \\
=\frac 12\mu{\rm sgn}(n-\mu)\pi-\xi_n(pr_c\hbar^{-1})-\frac 14\pi\label{eq65}
\end{eqnarray}
is the WKB phase shift, $\xi_n(s)$ and $\omega_n(s,\rho)$ are given
by (A.2) and (A.3) in Appendix A. The derivative of the WKB phase
shift, $\frac{\partial}{\partial n}\delta_n^{\rm WKB}$, can be
related to the classical reflection angle (equal to the classical
incidence angle), $\theta_{r_c}-\theta_\infty$, see (B.10) in
Appendix B:
\begin{equation}
\left.\left(\frac{\partial}{\partial n}\delta_n^{\rm WKB}\right)\right|_{n=\mu+\alpha/\hbar}=
\theta_{r_c}-\theta_\infty+\frac 12{\rm sgn}(\alpha)\pi.\label{eq66}
\end{equation}
Thus, the condition of stationarity of the phase in (64),
\begin{equation}
\left.\varphi=-2\left(\frac{\partial}{\partial n}\delta_n^{\rm WKB}\right)\right|_{n=n_0}\label{eq67}
\end{equation}
($\frac{\partial}{\partial n}\omega_n$ is negligible, see Appendix
A), coincides in its form with the classical relation between
scattering angle $\varphi$ and reflection angle
$\theta_{r_c}-\theta_\infty$, see (B.12) in Appendix B, and the
classical impact parameter can be introduced as
\begin{equation}
b=-\hbar(n_0-\mu)p^{-1};\label{eq68}
\end{equation}
note also that the first equation in (22) is equivalent to the
equation determining the squared $r$-derivative of the
Hamilton-Jacobi action, see (B.5) in Appendix B.

However, the merits of the WKB method are not restricted to
determining the WKB phase shift, i.e. to the only description of
classical elastic reflection. The use of the method has also allowed
us to find term (62) which is generically independent of the
self-adjoint extension parameter but depends on both the magnetic
flux and the particle momentum. This term describes the purely
quantum effect of diffraction. It should be noted that the quantum
effect is well separated in the scattering angle from the classical
one: namely, diffraction is in the forward direction where classical
reflection is absent. Although the range of angles where diffraction
is to be observed is quite narrow, the whole (i.e. integrated over
$-\pi<\varphi<\pi$) contribution of the diffraction effect to the
total cross section is the same as that of the reflection effect:
\begin{equation}
\frac{\rm d\sigma_1}{{\rm d} z}=\frac{\rm d\sigma_2}{{\rm d} z}=2r_c,\label{eq69}
\end{equation}
and the total cross section is twice the classical one; the latter
is needed for the optical theorem to be maintained (see
\cite{Si10,Si11}).

\section{Discussion and conclusion}

It should be noted in the first place that the AB scattering amplitude 
(see (51)) and, consequently, cross section (1) were obtained in 
\cite{Aha} by imposing the condition of regular behavior for the 
particle wave function at the location of a singular (i.e. of zero 
transverse size) magnetic vortex. However, the regularity condition 
is not the most general one that is required by probability 
conservation. The most general condition in the case when a spatial 
dimension along the vortex is ignored involves four self-adjoint 
extension parameters (the deficiency index is \{2,2\}) \cite{Ada,Dab}. 
The dependence on the self-adjoint extension parameters enters into 
terms which are added to amplitude $f_0$ (51). Therefore, the cross 
section depends on four arbitrary parameters or, if a longitudinal 
dimension is taken into account, on four arbitrary functions of $z$. 
If an invariance under rotations around the vortex is assumed to be 
a physical requirement, then the cross section still remains to be 
dependent on two arbitrary functions. The problem of seeking some, 
if any, sense for this arbitrariness is of purely academic interest, 
since the case of a magnetic vortex of zero transverse size has no 
relation to physics reality.

Turning now to the physical case of a magnetic vortex of nonzero 
transverse size, which is addressed in the present paper, we would
like to emphasize that the classical initial equation (see (B.2) in
Appendix B), as well as the quantum-mechanical one (see (10) with
(13)), involves the vector potential of the vortex (see (12)). However, 
in classical theory, this potential has no physical consequences, as 
is demonstrated, for example, in Appendix B. One may agree with the
authors of \cite{Bat} that ``the AB effect was already implicit in
the 1926 Schr\"{o}dinger equation'', but it takes more than
eight decades before a physical consequence of the vector
potential from the Schr\"{o}dinger equation is at last pointed out,
see (62). This explains why a scattering experiment aiming at the 
verification of the AB effect is still lacking. The main motivation 
of the present work is to make such an experiment plausible from a 
theorist's point of view. That is why we have taken care to prevent 
from any objection of theoretical kind: we have employed the most 
general boundary condition that is required by probability 
conservation and have shown how to grasp the AB effect even with 
the use of the 1926 WKB method.  

The AB effect, as well as quantization of the flux trapped in
superconductors, the Van Hove singularities in the excitation
spectra of crystals etc, explicates an importance of topological
concepts for quantum physics \cite{Yang}. The AB phase,
$e\Phi/(\hbar c)$, which is acquired by the charged-particle wave
function after the particle has encircled a magnetic vortex is
certainly the same for the particles of different energies.
Nonetheless, a question always is how to observe this phase in
real experiments, and what are physical restrictions on the
particle energy in these experiments. For instance, interference
experiments to verify the AB effect involve electrons of energies
restricted to the range of 10-100 KeV \cite{Ola}. Interference is
immanently a wave-optical phenomenon which is indirectly described 
in quantum theory through reference to wave optics (a
quantum-mechanical particle possesses de Broglie's wavelength and
thus is subject to the laws of wave optics). Meanwhile, quantum
theory provides a direct description of scattering phenomena and, 
in particular, of the AB effect as a scattering event. A scattering 
experiment with particles of the wavelength exceeding the vortex 
radius is hardly realistic, and that is why we are considering a 
scattering event with particles of the wavelength which is less 
than the vortex radius.

By using the WKB method in the present paper, we obtain the
scattering differential cross section, ${\rm d}\sigma/({\rm
d}z{\rm d}\varphi)$, consisting of two parts: one (61) describes
the classical effect of elastic reflection from the vortex and
another one (62) describes the quantum effect of diffraction on
the vortex. The effects are equal to each other, when integrated over
the whole range of the scattering angle, see (69). It should be
noted that, although the amplitude yielding the classical effect
depends on the self-adjoint extension parameter ($\rho$),
the flux of the vortex ($\Phi$) and the particle momentum ($p$),
see (56), all the dependence disappears in the differential cross
section, see (61); certainly, this cross section can be obtained
with the use of purely classical methods (the Hamilton-Jacobi
equation of motion).

The amplitude yielding the quantum effect is generically independent
of the self-adjoint extension parameter, see (57)-(60). This signifies 
that diffraction (to be more precise, the Fraunhofer diffraction, 
i.e. the diffraction in almost parallel rays) is independent of 
the choice of different boundary conditions which are compatible 
with the conservation of probability. Owing to diffraction the AB 
effect persists in scattering of real particles, e.g., electrons 
of energies in the range from 10 eV to 100 KeV. The classical 
effect opens a window in the forward direction, where the AB 
effect is to be observed as a fringe shift in the diffraction
pattern exhibited by the scattering differential cross section, see
(62).

We present the window for the observation of the AB effect on Fig.
1, where the scattering differential cross section and the
scattering angle are normalized in such a way that the plot
remains actually the same for $50<pr_c(2\pi\hbar)^{-1}<\infty$;
recall that the present-day realistic restriction is
$10^4<pr_c(2\pi\hbar)^{-1}<\infty$. The classical effect given by
(61) is unobservable at $|\varphi|<10\pi\hbar(pr_c)^{-1}$, whereas
the diffraction oscillations given by (62) are unobservable on the
classical background at $2\arcsin\left[10\hbar/(\pi
pr_c)\right]^{1/3}<|\varphi|<\pi$; at
$10\pi\hbar(pr_c)^{-1}<|\varphi|<2\arcsin\left[10\hbar/(\pi
pr_c)\right]^{1/3}$, both effects are unobservable being
indistinguishable from the $|f_3|^2$-background which is of order
of $\left[\hbar/(pr_c)\right]^{1/3}$, see (A.22) in Appendix A. A
gate for particles propagating in the strictly forward
($\varphi=0$) direction is opened when the vortex flux equals an
integer times the London flux: more than 90\% of the diffraction
cross section is accumulated in a peak centred at $\varphi=0$ and
having width $2\pi\hbar/(pr_c)$, see the dashed line on Fig. 1.
The gate for the strictly forward propagation of particles is
closed when the vortex flux equals a half-integer times the London
flux: a dip at $\varphi=0$ is surrounded by two symmetric peaks
accumulating more than 85\% of the diffraction cross section (each
peak having width $2\pi\hbar/(pr_c)$), see the solid line on Fig.
1. Thus, the classically forbidden propagation of particles in the
strictly forward direction is controlled by altering the amount of
the vortex flux.
\begin{figure}
\includegraphics[width=390pt]{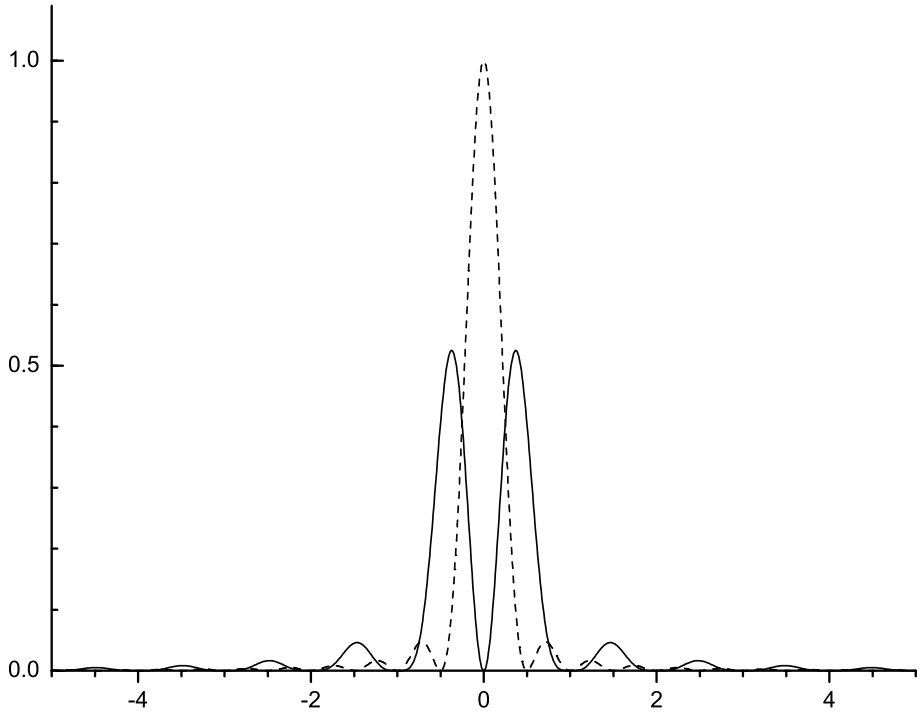}\\
\caption{$\frac{{\rm d}\sigma}{{\rm d}z{\rm
d}\varphi}/\left(\frac{pr_c}{h}\,\frac{{\rm d}\sigma}{{\rm
d}z}\right)$ is along the ordinate axis and $\varphi pr_c/h$ is
along the abscissa axis ($h=2\pi \hbar$). Dashed and solid lines
correspond to cases $\Phi=nhc/e$ and $\Phi=(n+1/2)hc/e$,
respectively ($n\in \mathbb{Z}$). The area under one central peak
by the dashed line ($|\varphi|<h/(2pr_c)$) is
0.4514119$\pm$0.0000002, and the area under two central peaks by
the solid line ($|\varphi|<h/(pr_c)$) is
0.4278549$\pm$0.0000013.}\label{1}
\end{figure}\normalsize

This scattering AB effect was never observed experimentally, since
a detailed proposal for its observation was just recently
elaborated \cite{Si12}. Such an experiment is in our opinion of
fundamental significance and seems to be quite feasible with the
present-day facilities providing for sources of bright and yet
coherent electron beams.

\section*{Acknowledgments}

The work of Yu.~A.~S. was supported by the National Academy of 
Sciences of Ukraine (project No.0112U000054), by the the Program 
of Fundamental Research of the Department of Physics and Astronomy 
of the National Academy of Sciences of Ukraine (project No.0112U000056) 
and by the ICTP -- SEENET-MTP grant PRJ-09 
``Strings and Cosmology''. The work of N.~D.~V. was supported by 
the Science and Technology Center in Ukraine and the National 
Academy of Sciences of Ukraine within the framework of the 
Targeted Research \& Development Initiatives (TRDI) Program 
under grant No.5716. We acknowledge the support from the State 
Agency for Science, Innovations and Informatization of Ukraine 
under the SFFR-BRRFR grant F54.1/019. 

\renewcommand{\thesection}{A}
\renewcommand{\theequation}{\thesection.\arabic{equation}}
\setcounter{section}{1} \setcounter{equation}{0}

\section*{Appendix A. Calculation of amplitudes $f_2$ and $f_3$}

We present amplitude $f_2$ (54) in the form
\begin{eqnarray}
f_2(p,\varphi)=-\frac{1}{\sqrt{2\pi p\hbar^{-1}}}\times \nonumber \\
\times\sum\limits_{|n-\mu|\leq pr_c/\hbar}
\exp\left\{{\rm i}\left[n\varphi+\mu\,{\rm sgn}(n-\mu)\pi-
2\xi_n(pr_c\hbar^{-1})-2\omega_n(pr_c\hbar^{-1},\rho)\right]\right\},
\end{eqnarray}
where, see (38),
\begin{equation}
\xi_n(s)=\frac 1\hbar\int\limits_{\hbar|n-\mu|/p}^{r_c}{\rm d}r\,P_n=\sqrt{s^2-(n-\mu)^2}-|n-\mu|
\arccos(|n-\mu|/s),
\end{equation}
\begin{equation}
\omega_n(s,\rho)=\arctan\left[\frac{\sqrt{s^2-(n-\mu)^2}}{{\rm cot}(\rho\pi)-
\frac 12\frac{s^2}{s^2-(n-\mu)^2}}\right],
\end{equation}
and we have introduced dimensionless variable $s=pr_c/\hbar$. To
calculate the sum in (A.1) in asymptotics $s\gg 1$, we use the
Poisson summation formula
\begin{equation}
\sum\limits_{|n-\mu|\leq s}e^{{\rm i}\eta(n,s)}=
\sum\limits_{l\in\mathbb{Z}}\int\limits_{-s_-}^{s_+}{\rm d}n
\exp\left\{{\rm i}[\eta(n,s)-2\pi nl]\right\}+\frac 12e^{{\rm i}\eta(s_+,s)}+ \frac 12e^{{\rm i}\eta(-s_-,s)},
\end{equation}
where $s_\pm=[\![s\pm\mu]\!]$. If $s_++s_-\gg 1$ and $\eta(n,s)$ is
convex upwards, $\frac{\partial^2\eta(n,s)}{\partial n^2}<0$, on
interval $-s_-<n<s_+$, then only a finite number of terms in the
series on the right-hand side of (A.4) contributes to the leading
asymptotics at $s\gg 1$, and one can use the method of stationary
phase for its evaluation. Namely, if equation
\begin{equation}
\left.\left[\frac{\partial}{\partial n}\eta(n,s)\right]\right|_{n=n_j}-2\pi l_j=0
\end{equation}
determines a stationary point inside the interval, $-s_-<n_j<s_+$ ,
for some values of $l$ denoted by $l_j$, then
\begin{equation}
\sum\limits_{|n-\mu|\leq s}e^{{\rm i}\eta(n,s)}=\sum\limits_{l_j}\exp
\left\{{\rm i}[\eta(n_j,s)-2\pi n_jl_j]\right\}\left\{\frac{2\pi e^{-{\rm i}\pi/2}}
{\left.-\left[\frac{\partial^2}{\partial n^2}\eta(n,s)\right]\right|_{n=n_j}}\right\}^{1/2}+O(1).
\end{equation}

In the present case we have
\begin{equation}
\eta(n,s)=n\varphi+\mu{\rm sgn}(n-\mu)\pi-2[\xi_n(s)+\omega_n(s,\rho)].
\end{equation}
Differentiating (A.2) and (A.3) over $n$, we obtain
\begin{equation}
\frac{\partial}{\partial n}\xi_n(s)=-{\rm sgn}(n-\mu)\arccos(|n-\mu|/s)
\end{equation}
and
\begin{equation}
\frac{\partial}{\partial n}\omega_n(s,\rho)=-\frac{\frac{n-\mu}{s}\left[1-\left(\frac{n-\mu}{s}\right)^2\right]^{1/2}
\left\{\left[1-\left(\frac{n-\mu}{s}\right)^2\right]\cot(\rho\pi)-\frac 32\right\}}
{\left\{\left[1-\left(\frac{n-\mu}{s}\right)^2\right]\cot(\rho\pi)-\frac 12\right\}^2+s^2\left[
1-\left(\frac{n-\mu}{s}\right)^2\right]^3}.
\end{equation}
(A.9) is negligible as compared to (A.8), being of order
$O(|n-\mu|/s^3)$ at $|n-\mu|\ll s$, $s\gg 1$, where the WKB
approximation is valid, see (25). Therefore equation (A.5) takes
form
\begin{equation}
\varphi+2{\rm sgn}(n_j-\mu)\arccos(|n_j-\mu|/s)-2\pi l_j=0.
\end{equation}
Choosing the range for $\varphi$ as $-\pi<\varphi<\pi$, we find that
a solution to (A.10) exists at $l_j=0$ only. Denoting the solution
by $n_j=n_0$, we obtain
\begin{equation}
n_0-\mu=-s\,{\rm sgn}(\varphi)\cos(\varphi/2),
\end{equation}
and, hence,
\begin{equation}
\left.\left[\frac{\partial^2}{\partial n^2}\eta(n,s)\right]\right|_{n=n_0}=-\frac
{2}{s|\sin(\varphi/2)|}.
\end{equation}
We see that the term corresponding to $l=0$ in the right-hand side
of (A.4) is of order $O(\sqrt{s})$, whereas all other terms are of
order $O(1)$. As a result, we obtain expression (56).

Amplitude $f_3$ (55) is presented in the form
\begin{eqnarray}
f_3(p,\varphi)=-\frac{1}{\sqrt{2\pi
p\hbar^{-1}}}\sum\limits_{|n-\mu|>s}
e^{{\rm i}[n\varphi+\mu{\rm sgn}(n-\mu)\pi]}\times \nonumber \\
\times\exp{\left[2\sqrt{(n-\mu)^2-s^2}-2|n-\mu|{\rm arccosh}(|n-\mu|/s)\right]}\times\nonumber \\
\times\left\{\frac{\cot(\rho\pi)+\frac 12\frac{s^2}{(n-\mu)^2-s^2}-\sqrt{(n-\mu)^2-s^2}}
{\cot(\rho\pi)+\frac 12\frac{s^2}{(n-\mu)^2-s^2}+\sqrt{(n-\mu)^2-s^2}}+\right.\nonumber \\\left.+
\frac{\rm i}2\exp\left[2\sqrt{(n-\mu)^2-s^2}-2(n-\mu){\rm arccosh}\left(\frac{|n-\mu|}{s}\right)\right]\right\}^{-1},
\end{eqnarray}
where (41) is taken into account. The sum in (A.13) at $\varphi=0$
is evaluated asymptotically at $s\gg1$ by converting it into
integral:
\begin{equation}
f_3(p,0)=-\sqrt{\frac{2r_c}{\pi}}\left\{\sqrt{s}\cos(\mu\pi)\left[I^{(1)}(s)-{\rm i}I^{(2)}(s)\right]
+O(1/\sqrt{s})\right\},
\end{equation}
where
\begin{eqnarray}
I^{(1)}(s)=4\int\limits_{1}^{\infty}{\rm d}v\exp\left[2s\left(v{\rm arccosh}v-\sqrt{v^2-1}\right)\right]
\frac{\cot(\rho\pi)+\frac 12\frac{1}{v^2-1}-s\sqrt{v^2-1}}{\cot(\rho\pi)
+\frac 12\frac{1}{v^2-1}+s\sqrt{v^2-1}}\times\nonumber \\
\times\left\{1+4\exp\left[4s\left(v{\rm arccosh}v-\sqrt{v^2-1}\right)\right]
\left[\frac{\cot(\rho\pi)+\frac 12\frac{1}{v^2-1}-s\sqrt{v^2-1}}{\cot(\rho\pi)+\frac 12
\frac{1}{v^2-1}+s\sqrt{v^2-1}}\right]^2\right\}^{-1}
\end{eqnarray}
and
\begin{eqnarray}
I^{(2)}(s)=2\int\limits_{1}^{\infty}{\rm d}v\times\nonumber \\ \times
\left\{1+4\exp\left[4s\left(v{\rm arccosh}v-\sqrt{v^2-1}\right)\right]
\left[\frac{\cot(\rho\pi)+\frac 12\frac{1}{v^2-1}-s\sqrt{v^2-1}}{\cot(\rho\pi)+\frac 12
\frac{1}{v^2-1}+s\sqrt{v^2-1}}\right]^2\right\}^{-1} .
\end{eqnarray}
In the cases of $|\cot(\rho\pi)|\gg s$ and $|\cot(\rho\pi)|\sim s$,
the integrand in (A.15) is maximal at $v=1$, and one can use the
Laplace method for the estimation of $I^{(1)}(s)$ in this case, see,
e.g., \cite{Erde}. Namely, let $g(v)$ be continuous function with
$0<g(1)<\infty$ and $h(v)$ be continuous differentiable function
obeying conditions $\frac{d}{dv}h(v)<0$ at $v>1$ and
$\frac{d}{dv}h(v)\approx-a(v-1)^{\nu-1}$ at $v\rightarrow 1$
$(\nu>0)$, then
\begin{equation}
\int\limits_{1}^{\infty}{\rm d}v g(v)e^{s\,h(v)}=\frac{g(1)}{\nu}e^{s\,h(1)}\Gamma\left(\frac 1\nu\right)
\left(\frac{\nu}{as}\right)^{1/\nu},
\end{equation}
where $\Gamma(y)$ is the Euler gamma-function. In the present case
$g(1)=1$, $h(1)=0$, $a=2\sqrt{2}$, $\nu=3/2$, and we obtain
\begin{equation}
I^{(1)}(s)=\Gamma\left(\frac{2}{3}\right)(12s^2)^{-1/3}.
\end{equation}
A similar estimate is obtained for $I^{(2)}(s)$ (A.16) in the case
of $|\cot(\rho\pi)|\gg s$:
\begin{equation}
I^{(2)}(s)=\frac 14\Gamma\left(\frac 23\right)\left(6s^2\right)^{-1/3}.
\end{equation}
In the case of $|\cot(\rho\pi)|\ll s$, the integrand in (A.16) is
maximal at $v=v_0$ where $v_0=1+2^{-5/3}s^{-2/3}$. The integrand in
the vicinity of the maximum is approximated as
$$
2\exp\left\{-9(2s)^{4/3}\exp[10(2s)^{-2/3}](v-v_0)^2\right\},
$$
and the integral is estimated as
\begin{equation}
I^{(2)}(s)=\frac{\sqrt{\pi}}{3}\left(\frac{2}{s^2}\right)^{1/3}.
\end{equation}
The integrand in (A.15) in the case of $|\cot(\rho\pi)|\ll s$
changes sign at $v=v_0$. Similarly to (A.20), we obtain the
estimate for integral $I^{(1)}(s)$ in this case:
\begin{equation}
I^{(1)}(s)=-\frac{1}{3}\left(\frac{1}{2s}\right)^{2/3}.
\end{equation}
In the case of $|\cot(\rho\pi)|\sim s$, the numerical analysis of
$I^{(2)}(s)$ yields an estimate of order $O(s^{-2/3})$. Hence, we
obtain the estimate for amplitude $f_3(p,\varphi)$ (A.13):
\begin{equation}
|f_3(p,\varphi)|\leq\sqrt{\frac{2r_c \, s}{\pi}} |I^{(1)}(s)-{\rm
i}I^{(2)}(s)|\leq\sqrt{r_c}as^{-1/6},
\end{equation}
where constant $a$ is independent of $s$, $\mu$ and $\varphi$.

\renewcommand{\thesection}{B}
\renewcommand{\theequation}{\thesection.\arabic{equation}}
\setcounter{section}{1} \setcounter{equation}{0}
\section*{Appendix B. Hamilton-Jacobi equation of motion and classical cross section}

In classical theory, we consider scattering of a point charged
particle by an impenetrable tube containing magnetic flux $\Phi$.
The Hamilton function corresponding to the motion outside of the
tube is
\begin{equation}
H=\frac{1}{2m}\left[p_r^2+\frac{1}{r^2}(p_\theta-\mu')^2+p_z^2\right],
\end{equation}
where $\mu'=e\Phi(2\pi c)^{-1}$, and the Hamilton-Jacobi equation
(see, e.g., \cite{New}) is
\begin{equation}
\frac{\partial S}{\partial t}+\frac{1}{2m}\left[\left(\frac{\partial S}{\partial r}\right)^2+
\frac{1}{r^2}\left(\frac{\partial S}{\partial \theta}-\mu'\right)^2+\left(\frac{\partial S}{\partial z}\right)^2\right]=0.
\end{equation}
A general solution to (B.2) is
\begin{equation}
S=-Ht+\int\limits^{r}{\rm d}r\,p_r+\int\limits^{\theta}{\rm d}\theta\,p_\theta+
\int\limits^{z}{\rm d}zp_z.
\end{equation}
Primary conservation laws are
\begin{equation}
p_\theta=\alpha,\qquad p_z=\tilde{\alpha},\qquad \frac{1}{2m}\left[p_r^2+\frac{1}{r^2}(p_\theta-\mu')^2+
p_z^2\right]=E
\end{equation}
with $\alpha$, $\tilde{\alpha}$ and $E$ being arbitrary constants
($E>0$). Taking the conservation laws into account, we obtain action
$S$ (B.3) in the form
\begin{equation}
S=-Et+\int\limits^{r}{\rm d}r\,\sqrt{2mE-\tilde{\alpha}^2-r^{-2}(\alpha-\mu')^2}+\alpha\theta+\tilde{\alpha}z+{\rm const}.
\end{equation}
Secondary conservation laws are
\begin{equation}
\frac{\partial S}{\partial \alpha}=\beta_\alpha,\qquad \frac{\partial S}{\partial\tilde{\alpha}}=\beta_{\tilde{\alpha}},
\qquad \frac{\partial S}{\partial E}=\beta_E
\end{equation}
with $\beta_\alpha$, $\beta_{\tilde{\alpha}}$ and $\beta_E$ being
arbitrary constants. Thus we obtain relations
\begin{equation}
\beta_\alpha=\theta-\int\limits^{r}{\rm d}r\frac{r^{-2}(\alpha-\mu')}
{\sqrt{2mE-\tilde{\alpha}^2-r^{-2}(\alpha-\mu')^2}},
\end{equation}
\begin{equation}
\beta_{\tilde{\alpha}}=z-\int\limits^{r}{\rm d}r\frac{\tilde{\alpha}}
{\sqrt{2mE-\tilde{\alpha}^2-r^{-2}(\alpha-\mu')^2}},
\end{equation}
\begin{equation}
\beta_E=-t+\int\limits^{r}{\rm d}r\frac{m}
{\sqrt{2mE-\tilde{\alpha}^2-r^{-2}(\alpha-\mu')^2}},
\end{equation}
which determine a trajectory of the particle, i.e. the functional
dependence of $\theta$, $z$ and $t$ on $r$. One can conclude that
the classical trajectory is independent of the enclosed magnetic
flux, since $\mu'$ can be absorbed into constant $\alpha$:
$\alpha-\mu'\rightarrow \alpha$. The trajectory is symmetric with
respect to the point of reflection, $r=r_c$.

As a consequence of (B.7), we obtain the following expression for
the incidence angle which is equal to the reflection angle:
\begin{equation}
\theta_{r_c}-\theta_\infty=\int\limits_{\infty}^{r_c}{\rm d}r\frac{r^{-2}\alpha}
{\sqrt{p^2-r^{-2}\alpha^2}}=-\arcsin\left(\frac{\alpha}{pr_c}\right),
\end{equation}
where $p=\sqrt{2mE-\tilde{\alpha}^2}$ $(2mE>\tilde{\alpha}^2)$.
Defining the impact parameter as
\begin{equation}
b=-\alpha p^{-1}\qquad (-r_c<b<r_c)
\end{equation}
and using elementary tools of plane geometry, we determine the angle
of deflection (called also the scattering angle) in range
$-\pi<\varphi<\pi$ as
\begin{equation}
\varphi={\rm sgn}(b)\pi-2(\theta_{r_c}-\theta_\infty)=2{\rm sgn}(b)\arccos(|b|/r_c).
\end{equation}
Hence, the scattering differential cross section is
\begin{equation}
\frac{{\rm d}\sigma^{\rm class}}{{\rm d}z{\rm d}\varphi}\equiv-\frac{{\rm d}b}{{\rm d}\varphi}=
\frac{r_c}{2}|\sin\frac{\varphi}{2}|.
\end{equation}


\begin{thebibliography}{99}

\bibitem{Ehre}%
W. Ehrenberg, R. E. Siday, Proc. Phys. Soc. London B \textbf{62}
(1949) 8.

\bibitem{Aha}%
Y. Aharonov, D. Bohm, Phys. Rev. \textbf{115} (1959) 485.

\bibitem{Cha}%
R. G. Chambers, Phys. Rev. Lett. {\bf 5} (1960) 3.

\bibitem{Wu}%
T. T. Wu, C. N. Yang, Phys. Rev. D {\bf 12} (1975) 3845.

\bibitem{Pes}%
M. Peshkin, Phys. Rep. {\bf 80} (1981) 375.

\bibitem{Rui}%
S. N. M. Ruijsenaars, Ann. Phys. (NY) \textbf{146} (1983) 1.

\bibitem{Au}%
Y. Aharonov, C. K. Au, E. C. Lerner, J. Q. Liang, Phys. Rev. D {\bf
29} (1984) 2396.

\bibitem{Ola}%
S. Olariu, I. I. Popescu, Rev. Mod. Phys. \textbf{57} (1985) 339.

\bibitem{Jac}%
R. Jackiw, Ann. Phys. (NY) \textbf{201} (1990) 83.

\bibitem{Be}%
M. V. Berry, J. Phys. A: Math. Theor. \textbf{43} (2010) 354002.

\bibitem{Ton}%
A. Tonomura, J. Phys. A: Math. Theor. \textbf{43} (2010) 354021.

\bibitem{Bat}
H. Batelaan, A. Tonomura, Phys. Today \textbf{62}, No.9 (2009) 38.

\bibitem{We}
G. Wentzel, Z. Phys. \textbf{38} (1926) 518.

\bibitem{Kra}
H. A. Kramers, Z. Phys. \textbf{39} (1926) 828.

\bibitem{Bri}
L. Brillouin, J. de Phys. \textbf{7} (1926) 353.

\bibitem{New}%
R. G. Newton, {\it Scattering Theory of Waves and Particles},
Springer-Verlag, Berlin, 1982.

\bibitem{Lan}%
L. D. Landau, E. M. Lifshits, {\it Course of Theoretical Physics
Vol.3 Quantum Mechanics. Non-relativistic Theory}, Pergamon Press,
Oxford, 1991.

\bibitem{Ree}%
M. Reed, B. Simon, {\it Methods of Modern Mathematical Physics II.
Fourier Analysis, Self-Adjointness}, Academic Press, New York, 1975.

\bibitem{Alb}%
S. Albeverio, F. Gesztezy, R.
Hoegh-Krohn, H. Holden, {\it Solvable Models in Quantum Mechanics},
Springer-Verlag, Berlin, 1988.

\bibitem{Wolf}%
M. Born, E. Wolf, {\it Principles of Optics}, Pergamon Press, Oxford,
1980.

\bibitem{SiV}%
Yu. A. Sitenko, N. D. Vlasii, J. Phys. A: Math. Theor.
 \textbf{44} (2011) 315301.

\bibitem{Vl}%
N. D. Vlasii, Problems of Atomic Science and Technology
\textbf{2012}, No.1 (2012) 93.

\bibitem{Si10}%
Yu. A. Sitenko, N. D. Vlasii, Europhys. Lett. \textbf{92} (2010)
60001.

\bibitem{Si11}
Yu. A. Sitenko, N. D. Vlasii, Ann. Phys. (NY) \textbf{326} (2011)
1441.

\bibitem{Ada}%
R. Adami, A. Teta, Lett. Math. Phys. \textbf{43} (1998) 43.

\bibitem{Dab}%
L. Dabrowski, P. Stovicek, J. Math. Phys. \textbf{39} (1998) 47.

\bibitem{Yang}
C. N. Yang, Intern. J. Mod. Phys. A \textbf{27} (2012) 1230035.

\bibitem{Si12}
Yu. A. Sitenko, N. D. Vlasii,  J. Phys. A: Math. Theor. {\bf 45}
(2012) 135305.

\bibitem{Erde}
A. Erdelyi, {\it Asymptotic Expansions}, Dover, New York, 1956.

\end{thebibliography}
\end{document}